\def\rfr#1{eq. (\ref{#1})}
\def\rfrs#1#2{(\ref{#1})-(\ref{#2})}
\def\leti{Lense--Thirring}

\def\derp#1#2{\rp{\partial{#1}}{\partial{#2}}}

\def\bm#1{{\mbox{\boldmath$#1$\unboldmath}}}
\def\bar{\begin{eqnarray}}
\def\ear{\end{eqnarray}}

\def\eqi{\begin{equation}}
\def\eqf{\end{equation}}
\def\eqia{\begin{eqnarray}}
\def\eqfa{\end{eqnarray}}
\def\rp#1#2{{#1\over#2}}

\def\lb#1{\label{#1}}

\def\oc2{$\mathcal{O}(c^{-2})$}

\documentclass[11pt]{article}
\usepackage{amsmath,amsthm,amscd,amssymb}
\usepackage{latexsym}
\usepackage{graphicx,epsfig}
\usepackage[authoryear]{natbib}

\begin{document}

\noindent{\bf \LARGE{An assessment of the measurement of the
Lense-Thirring effect in the Earth gravity field, in reply to:
``On the measurement of the Lense-Thirring effect using the nodes
of the LAGEOS satellites, in reply to ``On the reliability of the
so far performed tests for measuring the Lense-Thirring effect
with the LAGEOS satellites'' by L. Iorio,'' by I. Ciufolini and E.
Pavlis }}
\\
\\
\\
{L. Iorio}\\
{\it Viale Unit$\grave{a}$ di Italia 68, 70125\\Bari, Italy
\\tel./fax 0039 080 5443144
\\e-mail: lorenzo.iorio@libero.it}

\begin{abstract}
In this paper we reply to recent claims by Ciufolini and Pavlis
about certain aspects of the measurement of the general
relativistic Lense-Thirring effect in the gravitational field of
the Earth.  I) The proposal by such authors of using the existing
satellites endowed with some active mechanisms of compensation of
the non-gravitational perturbations as an alternative strategy to
improve the currently ongoing Lense-Thirring tests is unfeasible
because of the impact of the uncancelled even zonal harmonics of
the geopotential and of some time-dependent tidal perturbations.
II) It is shown that their criticisms about the possibility of
using the existing altimeter Jason-1 and laser-ranged Ajisai
satellites are groundless.
%
%
III) Ciufolini and Pavlis also claimed that we would have
explicitly proposed to use the mean anomaly of the LAGEOS
satellites in order to improve the accuracy of the Lense-Thirrring
tests. We prove that it is false. In regard to the mean anomaly of
the LAGEOS satellites, Ciufolini himself did use such an orbital
element in some previously published tests.  About the latest test
performed with the LAGEOS satellites, IV) we discuss the
cross-coupling between the inclination errors and the first even
zonal harmonic as another possible source of systematic error
affecting it with an additional $9\%$ bias. V) Finally, we stress
the weak points of the claims about the origin of the two-nodes
LAGEOS-LAGEOS II combination used in that test.

\end{abstract}

Keywords: Lense-Thirring effect; Earth gravity field; active
spacecraft; passive spacecraft; polar orbits; mean anomaly

\section{Introduction}
The post-Newtonian Lense-Thirring (LT) effect is one of the few
predictions of the Einsteinian General Theory of Relativity (GTR)
for which a direct, clean and undisputable test is not yet
available with a really satisfying precision ($\leq 1\%$).

According to Einstein, the action of the gravitational potential
$U$ of a given distribution of mass-energy on a particle moving
with velocity $\bm v$ is described by the coefficients
$g_{\mu\nu},\ \mu,\nu=0,1,2,3$, of the space-time metric tensor.
They are determined, in principle, by solving the fully non-linear
field equations of GTR for the considered mass-energy content.
These equations can be linearized in the weak-field ($U/c^2 <<1$,
where $c$ is the speed of light in vacuum) and slow-motion
($v/c<<1$) approximation (Mashhoon et al., 2001; Ruggiero and
Tartaglia 2002),
valid throughout the Solar System, and look like the equations of
the linear Maxwellian electromagnetism. Among other things, a
noncentral, Lorentz-like force \eqi\bm F_{\rm LT}=-2m\left(\rp{\bm
v}{c}\right)\times \bm B_g\lb{fgm}\eqf acts on a moving test
particle of mass $m$. It  is induced by the post-Newtonian
component $\bm B_g$ of the gravitational field in which the
particle moves. $\bm B_g$ is related to the mass currents of the
mass-energy distribution of the source and comes from the
off-diagonal components $g_{0i}, i=1,2,3$ of the metric tensor.
Thanks to such an analogy, the ensemble of the gravitational
effects induced by mass displacements is also named
gravitomagnetism. For a central rotating body of proper angular
momentum \bm L the gravitomagnetic field is\eqi\bm B_g=\rp{G[3\bm
r(\bm r\cdot \bm L )-r^2 \bm L]}{cr^5}.\lb{gmfield}\eqf

One of the consequences of \rfr{fgm} and \rfr{gmfield} is a
gravitational spin--orbit coupling. Indeed, if we consider the
orbital motion of a particle in the gravitational field of a
central spinning mass,  it turns out that
%
the longitude of the ascending node $\Omega$ and the argument of
pericentre $\omega$ of the orbit of the test particle are affected
by tiny secular precessions $\dot\Omega_{\rm LT}$,
$\dot\omega_{\rm LT}$ (Lense and Thirring, 1918; Barker and
O'Connell, 1974; Cugusi and Proverbio, 1978; Soffel, 1989; Ashby
and Allison, 1993; Iorio, 2001a)
%
\begin{equation}
\dot\Omega_{\rm LT} =\frac{2GL}{c^2 a^3(1-e^2)^{\frac{3}{2}}},\
\dot\omega_{\rm LT} =-\frac{6GL\cos i}{c^2
a^3(1-e^2)^{\frac{3}{2}}},\lb{LET}
\end{equation} where $a,\ e$ and $i$ are the semimajor axis, the eccentricity and the inclination, respectively, of
the orbit and $G$ is the Newtonian gravitational constant. Note
that in their original paper Lense and Thirring (1918)
used the longitude of pericentre $\varpi\equiv\Omega+\omega$.

The gravitomagnetic force may have strong consequences in many
astrophysical and astronomical scenarios involving, e.g.,
accreting disks around black holes
(Stella et al., 2003), gravitational lensing and time delay
(Sereno 2003; 2005a; 2005b). Unfortunately, in these contexts the
knowledge of the various competing effects is rather poor and
makes very difficult to reliably extract the true gravitomagnetic
signal from the noisy background. E.g., attempts to measure the LT
effect around black holes are often confounded by the complexities
of the dynamics of the hot gas in their accretion disks. On the
contrary, in the Solar System space environments the LT effect is
weaker but the various sources of systematic errors are relatively
well known and we have the possibility of using various artificial
and natural orbiters both to improve our knowledge of such biases
and to design suitable observables circumventing these problems,
at least to a certain extent.

Up to now, all the performed and ongoing tests of gravitomagnetism
were implemented in the weak-field and slow-motion scenarios of
the Earth, the Sun and Mars gravitational fields.

Concerning the terrestrial environment,  in April 2004 the GP-B
spacecraft
(Everitt et al., 2001) was launched. Its aim
is the measurement of another gravitomagnetic effect, i.e. the
precession of the spins
(Schiff, 1960) of four superconducting gyroscopes carried onboard
with an expected accuracy of $1\%$ or better. According to
Nordtvedt (2003),
the multi-decade analysis of the Moon's orbit by means of the
Lunar Laser Ranging (LLR) technique yields a comprehensive test of
the various parts of order $\mathcal{O}(c^{-2})$ of the
post-Newtonian equation of motion. The existence of the LT
signature as predicted by GTR would, then, be indirectly inferred
from the high accuracy of the lunar orbital reconstruction. Also
the radial motion of the LAGEOS satellite would yield another
indirect confirmation of the existence of the LT effect
(Nordtvedt, 1988).
Recently, a test of the LT effect on the orbit of a test particle
was performed by Ciufolini and Pavlis (2004).
They analyzed the data of the laser-ranged LAGEOS and LAGEOS II
satellites in the gravitational field of the Earth by using an
observable proposed in (Ries et al., 2003a; 2003b; Iorio and
Morea, 2004; Iorio, 2005a; 2006a).
The total accuracy claimed by Ciufolini and Pavlis is 5-10$\%$ at
1-3 sigma, respectively, but such estimate is controversial
(Iorio, 2005b; 2006b; Lucchesi 2005)
for various reasons mainly related to a realistic assessment of
the impact of  the static and time-varying parts of the Earth
gravity field. The total error may be as large as about $20\%$ at
1 sigma level (Iorio, 2005a; 2006b). Proposals for the use of at
least one new satellite can be found, e.g., in (Ries et al.,
1989);
in (Iorio, 2005c)
it is re-examined in the context of the recent advances in our
knowledge of the terrestrial gravitational field.

In regard to the solar gravitomagnetic field, recent progresses in
the orbit determination of the Solar System planets (Pitjeva,
2005)
allowed, for the first time, to search for evidence of the LT
precessions of the planetary orbits (Iorio, 2005d; 2005e).
In fact, the general relativistic predictions are compatible with
the determined extra-perihelion advances of the inner planets, but
the errors are still large: a zero-effect hypothesis is compatible
with them as well, although at a worse level. However, the
continuous processing of accurate radar-ranging data and the
ongoing/planned spacecraft-based missions to Mercury, Venus and
Mars like, e.g., BepiColombo, Messenger and Venus Express should
improve the robustness and the precision of such preliminary
tests.

Very recently, a $6\%$ LT test on the orbit of the Mars Global
Surveyor (MGS) spacecraft in the gravitational field of Mars has
been reported (Iorio 2006c); indeed, the predictions of general
relativity are able to accommodate, on average, about $94\%$ of
the measured residuals in the out-of-plane part of the MGS orbit
over 5 years.

%
At present, there are no observational evidences pointing against
the existence of the gravitomagnetic force as predicted by general
relativity. Thus, the main interest in an evaluation of the
accuracy of our knowledge of the LT effect as much accurate and
reliable as possible mainly resides in estimating the quality of
orbitography and of the gravitational and non-gravitational force
modelling without neglecting gravitomagnetism.

The focus of this paper is on some issues related to the LT test
recently performed in the terrestrial gravitational field with the
nodes of LAGEOS and LAGEOS II and on possible alternative
strategies to enhance its reliability and accuracy by using some
of the existing active or passive satellites. The subsequent
debate has raised controversies often pursued by Ciufolini and
Pavlis (2005)
in a way completely inappropriate for scientific
works, both for content and language.

The main points  are
\begin{itemize}
  \item In (Ciufolini and Pavlis, 2005) it is written: ``[...] one wonders
 why the author of Iorio\footnote{It is the reference (Iorio, 2005b) of this paper.} (2005)
 has not proposed the use of the GRACE satellites which not only have by far
 better shape and orbital stability compared to JASON but they also carry ultra precise accelerometers
 that measure all non--conservative accelerations, so that they are in
 practice ``free-falling'' particles in vacuum, at least
 to the extent that is covered by the accuracy of these instruments.''

 The answer is
that the existing spacecraft with active mechanisms of
compensation of the non-gravitational forces (CHAMP, GRACE, GP-B)
fly at too low altitudes in polar orbits, so that the use of their
nodes, combined or not with those of the LAGEOS satellites, would
greatly enhance the systematic errors both due to the uncancelled
even zonal harmonic coefficients $J_{\ell}$ of the multipolar
expansion of the terrestrial gravitational potential and to
certain time-dependent tidal perturbations. This topic is
discussed in detail in Section \ref{polari}.
  \item In Section \ref{jason} we deal with the proposal of using
  also the data from the existing altimeter Jason-1 and laser-ranged Ajisai
  satellites (Iorio and Doornbos, 2005; Vespe and Rutigliano, 2005)
  to improve and enforce the accuracy and the
  reliability of the LT tests. Such a suggestion was only superficially criticized by
  Ciufolini and Pavlis (2005), without even quoting the
  relevant works.
  \item The title of Section 5 of (Ciufolini and Pavlis, 2005) is: ``On the use of
the mean anomaly and on the use of Jason to measure the
Lense-Thirring effect proposed in Iorio (2004) ''. Moreover, in
(Ciufolini and Pavlis, 2005) it is also written that ``[...] one
of the most profound mistakes and misunderstandings of
Iorio\footnote{It is the reference (Iorio, 2005b) of this paper.}
(2005) is the proposed use of the mean anomaly of a satellite to
measure the Lense-Thirring effect (in some previous paper by the
same author the use of the mean anomaly was also explicitly
proposed).'' Such claims are not correct, as it will be shown in
Section \ref{anomalia}.
\item
In Section \ref{inclinaz} we discuss the bias induced on the test
performed with the LAGEOS satellites by the cross-coupling between
the first even zonal harmonic of the geopotential and the errors
in the inclinations of LAGEOS and LAGEOS II. It yields a further
$9\%$ systematic bias.

\item
Finally, in Section \ref{plagio} we discuss the plagiarism by
Ciufolini and Pavlis (2004) about the combination of the nodes of
LAGEOS and LAGEOS II (Ries et al., 2003a; 2003b; Iorio and Morea,
2004; Iorio, 2005a; 2006a)
 used by them and
show that their claims about its origin are not exact.
\end{itemize}

\section{On the possible use of other existing active and passive spacecraft}\lb{polari}
\subsection{The linear combination approach: a brief review}\lb{ammazzo}
A general description of the approach involving the reduction of
the systematic error due to the mismodelling in the classical part
of the terrestrial gravitational field by using suitable linear
combinations of the satellites' nodes can be found, e.g., in
(Iorio, 2002a; 2004; 2005b)
and references therein; here we briefly recall it (see also
Section \ref{inclinaz}).

Among the six Keplerian orbital elements in terms of which it is
possible to parameterize the orbital motion of a test particle in
the gravitational field of a central body of mass $M$ (Roy, 2003),
the longitude of the ascending node $\Omega$, the argument of the
pericentre $\omega$ and the mean anomaly $\mathcal{M}$ undergo
secular precessions due to the even zonal harmonics $J_{\ell},\
\ell=2,4,6...$ of the multipolar expansion of the Newtonian part
of the gravitational potential of $M$. The general relativistic
gravitomagnetic force only affects $\Omega$ and $\omega$ with the
secular precessions of the Lense-Thirring effect. In principle,
$\Omega, \omega$ and $\mathcal{M}$ could all be used in order to
design suitable linear combinations in order to cancel out the
low-degree even zonal harmonics whose classical precessions are
much larger than the Lense-Thirring signals of interest. Such
combinations are obtained by explicitly writing down the
expressions of the residuals of $N$ orbital elements in terms of
$N-1$ even zonal harmonics and the Lense-Thirring effect,
considered as an entirely unmodelled feature of motion, and
solving them with respect to the latter. The so obtained
coefficients weighing the satellites' orbital elements depend on
their semimajor axes $a$, eccentricities $e$ and inclinations $i$:
they allow to cancel out the impact of the $N-1$ even zonal
harmonics considered. However, the mean anomaly cannot be used at
all because it is sensitive to huge non-gravitational
perturbations which especially affect the Keplerian mean motion
$n=\sqrt{GM/a^3}$ via the indirect effects on the semi-major axis
$a$. They have quadratic signature in time. Thus, only $\Omega$
and $\omega$ have been used in the so-far performed tests with the
LAGEOS satellites (Ciufolini et al., 1997; 1998; Ciufolini and
Pavlis, 2004).
%
\subsection{The use of the node of GP-B}
The possible use of the node of the drag-free spacecraft GP-B has
been considered in (Peterson, 1997; Iorio, 2005f).
The major problems come, in this case, from the fact that a nearly
perfect polar orbital configuration would make the node to precess
at a very low rate; for GP-B it amounts to 1136 years (Iorio,
2005f). This is a problem because one of the major time-dependent
tidal perturbation, i.e. the solar $K_1$ tide, which is not
cancelled out by the linear combination approach because it is a
tesseral ($m=1$) perturbation (Iorio, 2001b), has just the same
period of the satellite's node. So, over an observational time
span of about one year--which is the lifetime of GP-B with the
drag-free apparatus nominally working--it would resemble an
aliasing superimposed secular trend (Iorio, 2002b). For example,
in (Iorio, 2005f) a combination with the node of LAGEOS and GP-B
has been considered: the coefficient of GP-B would amount to -398,
thus fatally enhancing such tidal bias.
\subsection{The use
of the nodes of the other existing geodetic satellites and of
CHAMP and GRACE} Analogous conclusions can also be drawn for a
possible use of CHAMP and GRACE. Indeed, let us consider, e.g., a
combination with the nodes of LAGEOS, LAGEOS II, Ajisai,
Starlette, Stella, CHAMP and one of the two spacecraft of the
GRACE mission in order to cancel out the first six even zonal
harmonics $J_{\ell},\ell=2,4,6,8,10,12$. Apart from the
coefficient of LAGEOS, which is 1, the coefficients which weigh
the nodes of the other satellites are
\begin{eqnarray}
c_{\rm LAGEOS\ II}&=&0.3237,\\
c_{\rm Ajisai}&=&-0.0228,\\
c_{\rm Starlette}&=&0.0234,\\
c_{\rm Stella}&=&-0.1814,\\
c_{\rm CHAMP}&=&-12.377,\lb{ccha}\\
c_{\rm GRACE}&=&33.5348\lb{cgra}.
\end{eqnarray}
The slope of the gravitomagnetic trend is 3714.58 milliarcseconds
per year (mas yr$^{-1}$ in the following). The systematic error
due to the uncancelled even zonal harmonics $J_{14},\ J_{\rm 16},\
J_{18},...$ amounts to $4.4\%$, at 1-sigma level, according to a
calculation up to degree $\ell=42$ with the variance matrix of the
Earth gravity model EIGEN-CG01C (Reigber et al. 2006) which
combines data from CHAMP, GRACE and ground-based measurements.
Such error level is not competitive with those which can be
reached by the used combination with LAGEOS and LAGEOS II  and the
proposed combination involving also Ajisai and Jason-1 (Iorio and
Doornbos, 2005; Vespe and Rutigliano, 2005).
Indeed, EIGEN-CG01C yields for them a systematic error of
gravitational origin of about $6\%$ and less than  $2\%$,
respectively, at 1-sigma level. Moreover, while the error due to
the geopotential is sensitive to the even zonals, at most, up to
degree $\ell=20$, for such combinations,  the multi-combination
with CHAMP and GRACE is sensitive to a much larger number of even
zonal harmonics due to the inclusion of the lower altitude
satellites Starlette and, especially, Stella, CHAMP and GRACE.
This makes rather difficult and unreliable the evaluation of the
systematic bias induced by the static part of the geopotential
because the calculation of the coefficients $\dot\Omega_{.\ell}$
of the classical secular precessions (Iorio, 2003) yield unstable
results after degree $\ell\sim 40$. Thus, the $4.4\%$ estimate of
the bias due to the even zonals is probably optimistic. Another
serious drawback of such multi-combination is represented by the
relatively large magnitude of the coefficients \rfrs{ccha}{cgra}
which weigh the nodes of CHAMP and GRACE. Indeed, they enhance the
impact of all the uncancelled time-dependent perturbations among
which the solar $K_1$ tide is one of the most powerful.  The
periods of the related orbital perturbations amount to -2.63 years
for CHAMP and -7.23 years for GRACE. Such effects would represent
serious aliasing bias over the necessarily short observational
time span due to the limited lifetimes of CHAMP and GRACE with
respect to the geodetic satellites.
\subsection{The use of the nodes of the other existing geodetic
satellites} In regard to the possibility of only using the other
existing geodetic spherical satellites, mainly Ajisai, Stella and
Starlette due to their long data records available, this problem
has already been tackled in a number of papers, like (Casotto et
al., 1990; Iorio, 2002c; 2002d). The fact that the inclusion of
the other satellites in the linear combination scheme is still not
competitive, although the improvements due to the first models
from CHAMP and GRACE, was shown in (Iorio and Doornbos, 2005): the
systematic error of gravitational origin amounts to 31$\%$ with
EIGEN-CG01C.
\subsection{The impact of the new Earth gravity models from CHAMP and GRACE on the use of
a polar orbital geometry} The impact of the most recent Earth
gravity models from CHAMP and GRACE on the use of the node of a
single satellite in polar orbit has been discussed in detail in
(Iorio, 2005c). By using EIGEN-CG01C, it turns out that for a
semimajor axis of, e.g., 8000 km, quite larger than GP-B, CHAMP
and GRACE, the systematic error due to the full spectrum of the
even zonal harmonics would amount to $25\%$ for an inclination of
88 deg. In this case the period of the $K_1$ tide would amount to
$\sim 10^3$ days. Instead, for an inclination of 89.9 deg the bias
due to the even zonals would be 2$\%$ but the period of the tidal
perturbation would raise to $\sim 10^4$ days. Also with the new
terrestrial gravity field solutions the linear combination
approach would fail.
\section{The proposed use of Jason-1 and Ajisai}\lb{jason}
In a recent paper (Iorio and Doornobos, 2004), we  proposed to
investigate the possibility of analyzing a suitable linear
combination involving the nodes of\footnote{See also (Vespe and
Rutigliano, 2005) on the same topic. } LAGEOS, LAGEOS II, Ajisai
and Jason-1.
\subsection{Advantages}
The advantages of the combination involving also Jason-1 and
Ajisai are
\begin{itemize}
\item
Cancellation of the first three even zonal harmonics $J_2,\ J_4,\
J_6 $  of the geopotential along with their secular variations
$\dot J_2$, $\dot J_4$, $\dot J_6$. Moreover, the systematic error
due to the remaining higher degree even zonal harmonics $J_8,\
J_{10},...$ is almost model-independent: indeed, it is $\lesssim
2\%$ (1-sigma level) according to the 2nd generation GRACE-only
Earth gravity models EIGEN-GRACE02S (Reigber et al., 2005) $(2\%)$
and GGM02S $(2.7\%)$ and to the model EIGEN-CG01C (Reigber et al.,
2006) ($1.6\%$). It should be mentioned that it is expected that
GRACE will yield a better improvement in the knowledge of the
higher degree even zonal harmonics, to which the Jason's
combination is sensitive, instead of the lower degree even zonals,
which mainly affect the node-node LAGEOS-LAGEOS II combination.
This means that it may happen that, in the near future, the bias
due to the geopotential will reduce down to $\sim 1\%$ for the
Jason's combination in a satisfactorily model-independent way,
while it may remain more or less unchanged for the two-nodes
LAGEOS-LAGEOS II  combination. The latest results obtained with
the combined model EIGEN-CG03C (F\"{o}rste et al., 2005) seems to
confirm this trend (Iorio, 2006d).

Finally, only the first ten even zonal harmonics (i.e. up to
$J_{20}$) should be accounted for in the sense that the systematic
error due to the geopotential does not change with the inclusion
of more zonals in the calculation.
\item Small coefficient, $\sim 10^{-2}$, of the node of Jason-1.
This is particularly important for reducing the impact of the
non-gravitational acceleration suffered by Jason-1.
\item No secular or long-period perturbations of gravitational and
non-gravitational origin.
\end{itemize}
\subsection{Drawbacks}
The possible weak points of this proposals are mainly the
following
\begin{itemize}
\item The huge impact of the non-gravitational forces--mainly atmospheric
drag, direct solar radiation pressure and Earth's albedo--on
Jason-1 which has not a spherical shape, being endowed with solar
panels. Moreover, they should be modelled in a truly dynamical way
in order to avoid to absorb the Lense-Thirring effect as it would
happen in the empirical reduced-dynamic approach adopted so far.
\item The difficulty of getting a smooth long time series of its
orbit also due to the periodical orbital maneuvers. On the other
hand, it should be noted that, due to the Jason's main goal which
is ocean altimetry, only the radial and transverse components of
its orbit have received major attention up to now. Also the
orbital maneuvers are mainly, although not entirely, in the
orbital plane. Instead, the node is related to the out-of-plane
component of the orbit.
\end{itemize}
However, a detailed, although preliminary, evaluation of the
impact of the Jason's non-conservative forces on the entire
combination has been performed in (Iorio and Doornbos, 2005). No
secular effects occur. On the contrary, a relatively
high-frequency (the 120-days period of the $\beta^{'}$ cycle
related to the orientation of the orbital plane with respect to
the Sun) of non-gravitational origin has been found. Its impact on
the suggested measurement of the Lense-Thirring effect has been
evaluated to be $\leq 4\%$ over a 2-years time span (without
removing such periodic signal).

Ciufolini and Pavlis (2005) do not consider the content of (Iorio
and Doornbos, 2005) and do not even try to discuss it in depth.
Moreover, they neither present data analysis nor numerical
simulations to convincingly support their a-priori claimed
infeasibility of our proposal.
\section{On the use of the mean anomaly}\lb{anomalia}
In regard to the mean anomaly issue, the only work quoted by
Ciufolini and Pavlis (2005) is an old, unpublished version of the
preprint gr-qc/0412057 in which the mean anomaly is not mentioned
at all. The same holds also for all the other subsequent versions
of that preprint, some of which appeared well before the
submission of (Ciufolini and Pavlis, 2005). Moreover, no other
papers by us in which such an alleged explicit proposal would
appear are explicitly cited in (Ciufolini and Pavlis, 2005). In
fact, in the whole literature does not exist any published paper
by us in which we explicitly put forth the possibility of using
the mean anomaly of the LAGEOS satellites for measuring the
Lense-Thirring effect. On the contrary, the use of the node and
the perigee of the Earth' satellites is always explicitly
mentioned in many papers where the linear approach combination is
exposed and generalized to other situations (Iorio, 2002a; 2002c;
2002d; 2003; 2004; 2006a).
%
%
%
Moreover, in (Iorio, 2005b), after a general description of
 the linear combination approach for
the measurement of the Lense-Thirring effect, it is unambiguously
written ``In general, the orbital elements employed are the nodes
and the perigees and the even zonal harmonics cancelled are the
first N-1 low-degree ones.'' Ciufolini and Pavlis (2005) do not
quote such sentence. In their paper it is written: ``A more
detailed work discussing [...] the use of the mean anomaly and
other highly unfeasible proposals by Iorio (2005) will be the
subjects of following paper''. It is, thus, likely that the
authors of (Ciufolini and Pavlis, 2005), after a careful
inspection of the relevant literature, will certainly experience
many difficulties in finding material for such an announced paper,
at least as far as the use of the mean anomaly for measuring the
Lense-Thirring effect is concerned.


Instead, it turns out that one of the authors of (Ciufolini and
Pavlis, 2005) used the semimajor axis, the eccentricity and the
mean anomaly of LAGEOS II in some (not explicitly released) linear
combination with the nodes of LAGEOS and LAGEOS II and the perigee
of LAGEOS II in some previous tests with the EGM96 Earth gravity
model (Figure 4 and pp. 47-48 of (Ciufolini, 2001); Figure 4 and
paragraph (b), pp. 2376-2377 of (Ciufolini, 2000)).
\section{The impact of the errors in the inclination on the latest test performed with the LAGEOS
satellites}\lb{inclinaz}
The adopted observable is the following combination of the
residuals $\delta\dot\Omega_{\rm obs}$ of the rates of the nodes
of LAGEOS and LAGEOS II (Iorio and Morea, 2004; Iorio, 2005a;
2006a) \eqi\delta\dot\Omega_{\rm obs}^{\rm
LAGEOS}+c_1\delta\dot\Omega_{\rm obs}^{\rm LAGEOS\ II}\sim
\mu_{\rm LT }S_{\rm LT},\lb{iorform}\eqf where $c_1\sim 0.546$,
$S_{\rm LT}=48.1$ mas yr$^{-1}$ is the slope of the secular trend
according to general relativity, and $\mu_{\rm LT}$ is equal to 1
in Einsteinian theory and 0 in Newtonian mechanics. The idea of
using only the nodes of the LAGEOS satellites to de-couple the
Lense-Thirring effect from $J_2$ was proposed in (Ries et al.,
2003a; 2003b) in the context of the expected improvements in our
knowledge of the Earth's gravity field from the dedicated GRACE
mission. The linear combination approach (see also Section
\ref{ammazzo}) was adopted for the first time by Ciufolini (1996)
for his early, less precise tests with the nodes of the LAGEOS
satellites and the perigee of LAGEOS II (Ciufolini et al., 1997;
1998). The combination of \rfr{iorform} has been built up in order
to cancel out the first even zonal harmonic coefficient $J_2$ of
the multipolar expansion of the Newtonian part of the terrestrial
gravitational potential, along with its time-varying part which
also includes its secular variations $\dot J_2$. Up to now, the
controversy about the total error budget was mainly focused on the
reliable evaluation of the impact of the remaining uncancelled
higher-degree even zonal harmonics $J_4, J_6,...$ along with their
secular variations $\dot J_4, \dot J_6$ (Ciufolini and Pavlis
2005; Iorio 2005b; 2006b; Lucchesi 2005).

Another source of systematic error which was always explicitly
neglected in the previously cited works is given by the
cross-coupling between the classical precessions due to the even
zonal harmonics and the uncertainties in our knowledge of the
satellites' inclinations (Ciufolini et al. 1997). For the sake of
clarity, let us focus on $J_2$: the classical node precession
induced by it is
\eqi\dot\Omega_{J_2}=-\rp{3}{2}\left(\rp{R}{a}\right)^2\rp{n\cos i
 J_2}{(1-e^2)^2},\eqf
where $n=\sqrt{GM/a^3}$ is the Keplerian mean motion and $R$ is
the Earth's mean equatorial radius. The node orbital residuals
$\delta\dot\Omega_{\rm obs}$ account not only for the mismodelling
in $J_2$, whose cancellation is the goal of \rfr{iorform}, but
also for the uncertainty in $i$ which yields
\eqi\delta\dot\Omega_{i}\equiv\derp{\dot\Omega_{J_2}}{i}\delta i=
\rp{3}{2}\left(\rp{R}{a}\right)^2\rp{n\sin i
 J_2}{(1-e^2)^2}\delta i.\lb{nodii}\eqf
 The mismodelled precessions of \rfr{nodii} are not cancelled out
 by the combination of \rfr{iorform}. The total error is, thus, conservatively evaluated as
 \eqi\delta\mu_i\leq\left|\delta\dot\Omega_{i}^{\rm LAGEOS}\right|+
 \left|c_1\delta\dot\Omega_{i}^{\rm LAGEOS\ II}\right|\lb{ERRI}.\eqf
 An averaged root-mean-square error of 0.5 mas for both the LAGEOS satellites has been obtained by processing
 their data with a software like GEODYN; thus, a bias of 4.5 mas yr$^{-1}$ is
 obtained from \rfr{ERRI} at 1-sigma level. It amounts to about\footnote{If, instead, a
 root-sum-square calculation is performed, $\delta\mu_i=
 \sqrt{\left(\delta\dot\Omega_{i}^{\rm LAGEOS}\right)^2+
 \left(c_1\delta\dot\Omega_{i}^{\rm LAGEOS\ II}\right)^2}$ amounts to 6$\%$ of the Lense-Thirring effect.} $9\%$ of the Lense-Thirring
 effect.

 This further contribution to  the total error budget must be added
 to all
 the previous estimates which accounted only for the impact of $\delta\dot J_{\ell}$ and/or $\delta J_{\ell}$,
 $\ell=4,6,...$ (Ciufolini and Pavlis 2005; Iorio 2005b; 2006b; Lucchesi
 2005). In the most favorable case, i.e. a $\sim$10$\%$ gravitational
 bias mainly due to $\delta J_{\ell}$ only (Ciufolini and Pavlis 2005; Lucchesi 2005),
 the total systematic error becomes, thus,
 19$\%$. If, instead, the estimates by Iorio (2006b) of the gravitational error are
 used, which emphasize the impact of $\delta\dot J_{\ell}$,
 the total systematic error becomes close to 30$\%$.

\section{On the combination of the nodes of LAGEOS and LAGEOS II}\lb{plagio}
In (Ciufolini and Pavlis, 2005) it is claimed that one of the
authors would have put forth the idea of suitably combining the
Keplerian orbital elements of the LAGEOS-type satellites in order
to disentangle the Lense-Thirring effect from some of the much
larger biasing secular precessions induced by the even zonal
harmonics of the geopotential since the beginning of his studies
on the measurability of the gravitomagnetic effect, when the
LAGEOS II, launched in 1992, did not yet exist. This claim is
supported by quoting from (Ciufolini, 1986; 1989):
 ``[...] A solution would be to orbit several
high-altitude, laser-ranged satellites, similar to LAGEOS, to
measure $J_2,$ $J_4$, $J_6$, etc., and one satellite to measure
$\dot{\it \Omega}_{\rm Lense-Thirring }.$''  It seems more likely
that this statement could, at most, mean that one should first use
many satellites to accurately measure the various Earth's even
zonal harmonics with a sufficiently high accuracy and, then,
analyze the node only of one satellite for safely measuring its
gravitomagnetic Lense-Thirring precession. However, there is no
trace at all of the linear combination approach which will be
introduced only in 1996 (Ciufolini, 1996) in the particular case
of the nodes of LAGEOS and LAGEOS II and the perigee of LAGEOS II.
Indeed, the previously quoted statement comes after a discussion
of the impact of $J_2$ and of the higher degree even zonal
harmonics on the possible use of the node only of LAGEOS;
moreover, the rest of the papers (Ciufolini, 1986; 1989)
deals with the supplementary configuration
LAGEOS-LAGEOS X (later LAGEOS-LAGEOS III/LARES/WEBER-SAT).

The remarks about the triviality of the simple algebraic linear
system of two equations in two unknowns with which the
LAGEOS-LAGEOS II combination is obtained seem to be  a little
inappropriate because they could also be extended to the original
node-node-perigee combination (Ciufolini, 1996)
and the related linear algebraic system of three equations in
three unknowns.

In regard to the combination of \rfr{iorform}, our papers on it
are available on the Internet since April 2003. Moreover, we
personally know the authors of (Ciufolini and Pavlis, 2005) having
collaborated with them for some years and sent them various
e-mails between April and September 2003 with our results
attached; the interested reader can ask us for them. In one case
(28 March 2003 and 30 March 2003), Ciufolini asked us to prepare a
short table with our results in .doc format in view of a
video-conference with NASA'officials to be attended in the
following days by Ciufolini. A few days before, 26 March 2003, we
e-mailed  the .pdf of (Iorio and Morea, 2004) to Ciufolini. Later,
on 7 September 2003, we discussed the impact of the GGM01C Earth
gravity model on the LAGEOS-LAGEOS II combination with Pavlis.

Thus, the style and the content of the following comment present
in (Ciufolini and Pavlis, 2005): ``In conclusion, all the claims
of Iorio (2005) are simply lacking of any rational basis: above is
shown how much work was already published on this topic before
Iorio (2005) even began to produce any of his paper on this topic
and to rediscover some earlier results [...]. To avoid the
misunderstandings of Iorio (2005), it would have just been a
matter of very careful reading the previously existing literature
on this subject!'' are clearly inappropriate.
\section{Conclusions}
In this paper we replied to certain criticisms by Ciufolini and
Pavlis related to some issues concerning possible alternative
strategies aimed to improve the quality of the currently ongoing
Lense-Thirring tests with the LAGEOS satellites by using other
existing active or passive spacecraft. Indeed, the combination of
the nodes of LAGEOS and LAGEOS II satellites recently used
represents undoubtedly a notable improvement with respect to a
previous one including also the perigee of LAGEOS II. By the way,
it is mainly sensitive to $J_4, J_6$ and $\dot J_4,\dot J_6$;
unless the knowledge of such part of the terrestrial gravitational
field will notably be improved, the precision of the
Lense-Thirring tests performed with such two-elements combination
cannot be enhanced. A further  source of systematic bias is
represented by the uncertainties in the inclinations of the
satellites. A thorough discussion about the realistically
obtainable accuracy in our knowledge of such a gravitomagnetic
feature is motivated not by some existing discrepancies among the
general relativistic predictions for it and the so far collected
and  analyzed observations, but by the need of improving precise
orbitography and dynamical modelling without neglecting the
Lense-Thirring effect itself.

I) We demonstrated that the recent proposal by Ciufolini and
Pavlis of using the orbital data from the existing terrestrial
spacecraft endowed with some active mechanism of compensation of
the non-gravitational perturbations like GP-B, CHAMP and GRACE is
not suitable for measuring the Lense-Thirring effect both because
of their too low altitude and of their nearly polar orbital
geometry. Indeed, they would introduce in the resulting linear
combinations much more even zonal harmonics of high degree (more
than $\ell=40$) which would enhance the induced systematic error
and make unreliable its calculation. Moreover, the polar geometry
of GP-B, CHAMP and GRACE would have, as a consequence, that the
coefficients with which they would enter the combinations would be
larger than 1, thus enhancing various uncancelled time-dependent
long-period perturbations like that due to the $K_1$ tide. The new
Earth gravity models from CHAMP and GRACE do not alter this
situation.

II) The only possibility of enhancing the reliability and the
accuracy of the Lense-Thirring tests, at least with the existing
spacecraft, is linked to the feasibility of the data analysis of a
combination including also Jason-1 and Ajisai, although it will
not be a trivial task due to the impact of the non-gravitational
perturbations on Ajisai and, especially, Jason-1.

III) The claims by Ciufolini and Pavlis about an alleged proposal
by us of using the mean anomaly of the laser-ranged satellites to
test the Lense-Thirring effect have been proven to be groundless.

IV) The cross-coupling between $J_2$ and the errors in the
satellites' inclinations lead to a further systematic bias which
cannot be removed by the combination of the nodes of LAGEOS and
LAGEOS II. Its impact on the measurement amounts to 9$\%$, a
figure which must be added to all the previous estimates of the
total error budget which only account for the  even zonal
harmonics and their temporal variations.

V) Finally, we have shown that the arguments by Ciufolini and
Pavlis about the origin of the node-node LAGEOS-LAGEOS II
combination are not correct.
\section*{Acknowledgements}
I gratefully thank the anonymous referees for their efforts which
greatly contributed to significantly improve this paper.


\end{document}